# A New Clustering Approach For Anomaly Intrusion Detection

Ravi Ranjan and G. Sahoo

Department of Information Technology, Birla Institute of Technology, Mesra, Ranchi

*ABSTRACT*

*Recent advances in technology have made our work easier compare to earlier times. Computer network is growing day by day but while discussing about the security of computers and networks it has always been a major concerns for organizations varying from smaller to larger enterprises. It is true that organizations are aware of the possible threats and attacks so they always prepare for the safer side but due to some loopholes attackers are able to make attacks.*

*Intrusion detection is one of the major fields of research and researchers are trying to find new algorithms for detecting intrusions. Clustering techniques of data mining is an interested area of research for detecting possible intrusions and attacks. This paper presents a new clustering approach for anomaly intrusion detection by using the approach of K-medoids method of clustering and its certain modifications. The proposed algorithm is able to achieve high detection rate and overcomes the disadvantages of K-means algorithm.*

## 1. INTRODUCTION

The threat landscape is continuously changing day by day and new attacks are emerging in an unexpected manner. The large infrastructure of computer networks is often vulnerable towards the perils of the attacks and which has been disconcerting and embarrassing situation for the organizations which are new in the market. To cope up with these attacks James Anderson[1] first introduced the concept of intrusion detection in 1980 and a model based on this was introduced by Denning in 1987[2].

Intrusion Detection System (IDS) is a device typically another separate computer that monitors activity to identify malicious or suspicious events. An IDS is a sensor, like a smoke detector, that raises an alarm if specific things occur [3]. A model of IDS is shown in the figure 1.

The components in the figure are the four basic elements of an intrusion detection system, based on the common intrusion detection framework of [4]. An IDS receives raw inputs from sensors. It saves those inputs, analyzes them, and takes some controlling action. Till now many approaches have been proposed to resolve the problem of IDSs among which data mining techniques are most popular and successful. Clustering is the most important techniques of data mining which has been widely used and acceptable.

Clustering is an unsupervised method which takes a different approach by grouping objects into meaningful subclasses so that members from the same cluster are quite similar and different to the members of different cluster. If we use clustering for intrusion detection then we have in anomaly detection model which is developed based on normal data and deviations are searched over this model. It is difficult to say that there are no attacks during the time traffic collected from the network.

The unsupervised anomaly detection algorithm clusters the unlabeled data instances together into clusters using a simple distance-based metric. Once data is clustered all of the instance that appear in small clusters are labeled an anomalies because normal instances should form large clusters compared to intrusions and malicious intrusions and normal instance are qualitatively different so they do not fall into same cluster.

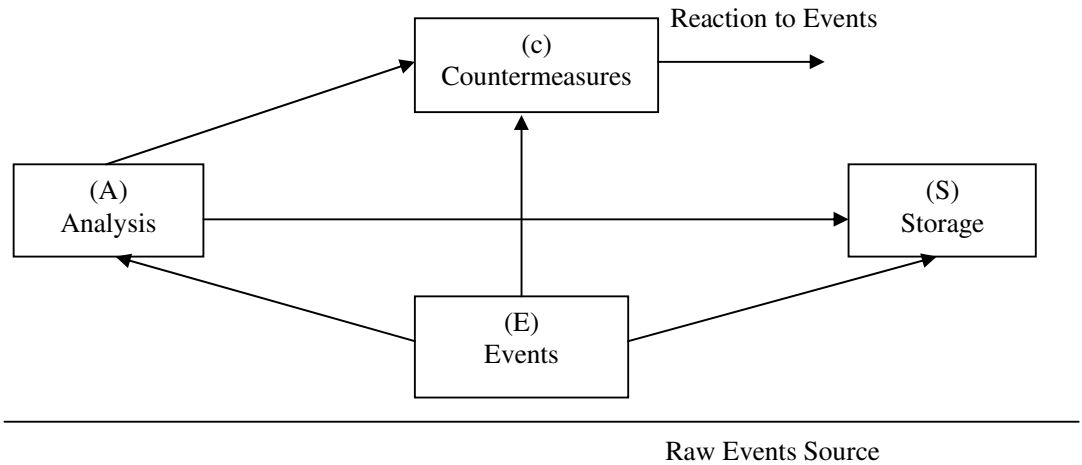

Figure 1. Common components of an Intrusion Detection Framework

## 2. RELATED WORK

Many works have been done and published if we discuss about clustering and data mining techniques. M. Jianliang, et al. [5] introduced the application on intrusion detection based on K-means clustering algorithm. K-means is used for intrusion detection to detect unknown attack and partition large data space effectively but it has many disadvantages degeneracy, cluster dependence .Yu Guan, et al. [6] introduced Y-means algorithm which is a clustering method of intrusion detection. This algorithm is based on K-means algorithm and other related clustering algorithm. It overcomes two short comings of K-means, no of cluster dependency and degeneracy. Zhou mingqiang, et al. [7] introduced a new concept of a graph based clustering algorithm for anomaly based clustering algorithm for anomaly intrusion detection. They used outlier detection method which is based on local deviation coefficient (LDCGB). Compared to other intrusion detection algorithm of clustering this algorithm is unnecessary to initial cluster number. Chitrakar R and Huang chuanhe [8] proposed a hybrid learning approach of combining k-medoids clustering and naive bayes classification. Because of the fact that k-medoids technique represents the real world scenario of data distribution the proposed algorithm will group the whole data into clusters more accurately than K-means such that it results in better classification. Yang Jian [9] proposed an improved intrusion detection model based on DBSCAN which illustrates the idea of generating a cluster using density method and merging small clusters. The paper describes a more reasonable density-based clustering algorithm for intrusion detection (IIDGB), using rational method to calculate the distance and design the method of parameter selection.

Li Xue-yong and Gao Guo [10] proposed a new intrusion detection based on improved DBSCAN which uses an improved density based cluster algorithm to improve the drawbacks of earlier approach as proposed in [9] by using more rational method for calculating the distance and process of merging cluster. Some of the drawbacks which are overcome are high false alarm rate, generation of small clusters after the experiment. Lei Li, et al. [11] introduced a novel rule-based intrusion detection system using data mining. They proposed an improvement over apriori algorithm by bringing the concept of length-decreasing support to detect intrusion. Association rules and sequence rules are the main technique of data mining. Z. Muda, et al. [12] published new work of Intrusion detection based on K-means clustering and OneR classification; they proposed an approach which combines the techniques of K-means and OneR classification. The main goal of paper is to utilize K-means algorithm and to split and group data into normal and attack instances. The algorithm partition the dataset into k clusters according to an initial value known as seed point into each cluster's centroids or cluster centers. The mean value of each data contained within each cluster is called centroids. Zhengjie Li, et al. [13] proposed anomaly intrusion detection method based on K-means clustering algorithm with particle swarm optimization. Particle swarm optimization (PSO) algorithm is an evolutionary computing technology which is based on swarm intelligence has good global search ability. The proposed algorithm has overcome falling into local minima and has relatively good overall convergence. K. Wankhade, et al. [14] gave an overview of intrusion detection which is based on data mining techniques. They discussed the various data mining techniques which can be applied on intrusion detection system for the effective identification of both known and unknown pattern of attack in order to develop a secure information system. H. Fatma and L. Mohamed [15] proposed a two stage technique to improve intrusion detection system based on data mining algorithm. They adopted a two stage technique in order to improve the accuracy of sensors. The first stage aim to generate meta-alerts through clustering and the second stage aims to reduce the rate of false alarms using a binary classification of the generated meta-alerts. For the first stage they used two alternatives, self-organizing map (SOM) with K-means algorithm and neural GAS with fuzzy c-means algorithm and for the second stage they used three approaches, SOM with K-means algorithm, support vector machine and decision trees. A.M. Chandrasekhar and K. Raghuveer [16] introduced a new concept of Intrusion detection technique by using K-means, fuzzy neural network and SVM classifiers. The proposed technique has four major steps: first one is to use K-means algorithm to generate different training subsets. Based on the obtained training subsets, different neuro-fuzzy models are trained. Then a vector for SVM classification is formed and lastly, classification using radial SVM is performed to detect intrusion has happened or not.

The different approaches discussed above are proposed for intrusion detection using various techniques of data mining or other algorithm. They have many advantages over earlier approach but they are not good in all respect. In this paper we present a new approach for anomaly intrusion detection by using new medoid algorithm of K-medoid and certain modifications of it. The rest of the paper is organized as follows: section three discuss about the existing K-means algorithm in brief, section four discuss our proposed work in detail , section five discuss about the experimental methodology and results for our proposed work and finally in the last section we discuss our contributions and future work in this field.

## 3. K-MEANS ALGORITHM

For the completeness of the paper we discuss the K-means algorithm in this section. K-means is an iterative clustering algorithm in which items are processed among set of clusters until the desired set is reached. K-means algorithm is like a squared error algorithm. Using K-means algorithm we achieve a high degree of similarity among elements and dissimilarity in different clusters.

Given a cluster $K_i$, let the set of items mapped to that cluster be $\{t_{i1}, t_{i2}, t_{im}\}$.

The cluster mean of $K_i = \{t_{i1}, t_{i2}, t_{im}\}$ is defined as

$$m_i = \frac{1}{m}\sum_{j=1}^{m} t_{ij}$$

This definition indicates each tuple has only one numeric value as opposed to a tuple with many attribute values. The following sum up the steps of K-means algorithm [17].

---

Algorithm K-means

---

*Input:* E= $\{t_1, t_2....t_n\}$
      c  //number of desired clusters
*Output:* C //Set of clusters
*Begin*
    Assign initial values for mean $m_1, m_2, ... m_k$;
    Repeat
      Assign each item $t_i$ to the cluster 'c' which has the closest mean.
      Calculate new means for each cluster.
    Until
      Convergence criteria are met.
*End*

---

K-means has been widely used algorithm however it has many disadvantages.

It includes dependence on initial centroids, dependence on number of clusters and degeneracy. To overcome these disadvantages we have proposed a new algorithm.

## 4. PROPOSED ALGORITHM

To overcome the disadvantages of K-means and to improve detection rate we have proposed a new clustering approach for anomaly intrusion detection. In this approach k-medoids algorithm and its modifications are used.

The k-medoids algorithm is also a partitioning technique of clusters that clusters the data sets of n objects into k clusters with apriori. It could be more robust to noise and outliers as compared to K-means since it minimize a sum of pairwise dissimilarities using a squared Euclidean distance.
Definition: A medoid of a finite dataset is a data point from this set whose average dissimilarity to all data points is minimal.

The algorithm is described as follows.

---

New Medoid Clustering Algorithm

---

*Input*: D dataset of n object
*Output:* Desired set of normal and abnormal clusters.
*Begin*

Step1: Standardize the dataset in order to make the feature value to appropriate range.
      This is done because features with greater value dominate the features with lesser value.

Step2: Select initial medoids and for that the formula of Euclidean distance for dissimilarity measure has been used. It is given as under:

$$dist_{ij} = \sqrt{\sum_{b=1}^{y}(z_{ib} - z_{jb})^2} \quad , i=1, 2,....x \text{ and } j=1, 2,...x$$

Let x objects having y variables classifies into c clusters.

    Compute

$$y_{ij} = dist_{ij} / \sum_{k=1}^{x} dist_{ik}$$

After finding $y_{ij}$ at each object and sorting them in ascending order, c objects are selected as the initial medoids having minimum value.

Step3: Associate each object to its closest medoid and calculate the optimal value as the sum of distances from all objects to their medoids.

Step4: Swap the current medoid in each cluster by the object which minimizes total distance to other objects in the cluster.

Step5: Again associate each object to the closest medoids and compute the new value as in step3. If the new value is same as previous one then stop the algorithm otherwise repeat step4.

*End*

The above algorithm will result in cluster formation and the next steps is to check for an empty cluster, if there is an empty cluster then remove the empty cluster by deleting them and hence this will eliminate degeneracy problem.

## 5. EXPERIMENTAL DISCUSSIONS

The following points discuss about the steps of an experiment conducted.

### 5.1. Input data set

The KDD cup99 data set used as an input in order to conduct our experiment and later to check the performance of the algorithm. The kddcup99 dataset is the most commonly used dataset for intrusion detection first given by Massachusetts Institute of Technology. This dataset contains 24 kinds of attacks that can be categorized as four types to be named as denial of service attack, user to root attack, remote to local attack and probe attack. It contains 41 features divided into 34 nominal and 7 numeric features [18].

### 5.2. Data standardization

Standardization of the dataset has been done so that it would be appropriate to be used by the proposed algorithm. The following steps show the standardization process:

    Step 1: Find the mean of each feature in the dataset using the equation

$$\text{Mean}_f = \frac{D_{1f} + D_{2f} + \cdots D_{nf}}{n}$$

    where,

        $D_{1f}, D_{2f},.....D_{nf}$ are n measuring values of each feature f.

    Step 2: Compute the standard deviation of the calculated mean using the equation

33

International Journal of Data Mining & Knowledge Management Process (IJDKP) Vol.4, No.2, March 2014

$$SD_f = \frac{1}{n}(|D_{1f} - \text{Mean}_f| + |D_{2f} - \text{Mean}_f| + \ldots |D_{nf} - \text{Mean}_f|)$$

where ,

$D_{1f}, D_{2f}, \ldots D_{nf}$ are n measuring values of each feature f.

Step 3: The standardized value are given as:

$$S_{if} = \frac{D_{if} - \text{Mean}_f}{SD_f}$$

## 5.3. Execution and discussion

After getting the standardized value the proposed algorithm is being run to get the desired cluster and later we choose the cluster to label them as normal or intrusions. The proposed algorithm is compared against the existing algorithm on the basis of parameters detection rates, accuracy and false alarm rate.

The different metrics used for the purpose of checking performance and observing experimental results are given as under:

1. Detection Rate (DR)
2. False Alarm Rate (FAR)
3. False Positive (FP)
4. False Negative (FN)
5. True Positive (TP)
6. True Negative (TN)

1. Detection rate (DR): Detection rate is defined as the number of intrusion instance detected by the system (True Positive) divided by the total number of intrusion instances present in the test dataset.
2. False Alarm Rate (FAR): False Alarm Rate is defined as the number of normal patterns classified as attacks (False Positive) divided by total number of normal patterns.
3. False Positive (FP): False Positive corresponds to an event signaling IDS to produce an alarm when no attack has taken place.
4. False Negative (FN): False Negative is referred as a failure of IDS to detect an actual attack.
5. True Positive (TP): True Positive corresponds to a legitimate attack which triggers an IDS to produce an alarm.
6. True Negative (TN): True Negative corresponds to a situation when no attack has taken place and no alarm is raised.

The Detection rate is measured using the following formula:

Total no of Intrusion Instance= TP+FP

$$\text{Detection Rate (DR)} = \frac{TP}{\text{TOTAL NO OF INTRUSION INSTANCE IN DATASET}}$$

34

Accuracy is measured by the following formula:

$$\text{Accuracy} = \frac{TP+TN}{TP+TN+FP+FN}$$

False alarm rate is given by the following formula:

$$\text{False Alarm Rate (FAR)} = \frac{FP}{FP+TN}$$

The Detection Rate, Accuracy and False Alarm Rate are calculated to check the performance of our proposed approach with other existing algorithms.

The figure 2 depicts the percentage of accuracy for different algorithms including our proposed approach which is calculated using the formula of accuracy given earlier in the paper. The accuracy of the proposed algorithm is 96.38% which is comparatively higher than the existing K-means, FCM, and Y-means algorithms.

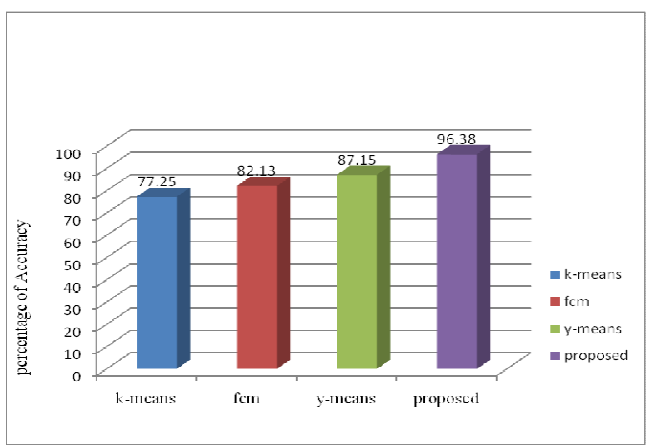

Figure 2. Accuracy comparison

Table 1. Detection Rate of Proposed Algorithm, K-means, FCM and Y-means

| Parameters | K-means | FCM | Y-means | Proposed algorithm |
|---|---|---|---|---|
| Detection Rate | 82.3 | 84.6 | 86.3 | 91.2 |
| Accuracy | 77.25 | 82.13 | 87.15 | 96.38 |
| False Alarm Rate | 5.2 | 4.2 | 3.9 | 3.2 |

The Table 1 gives the percentage of detection rate, accuracy and false alarm rate. The detection rate for our proposed method is 91.2 which is greater than K-means (82.3), Y-means (86.3) and FCM (84.6). Similarly the accuracy for our proposed algorithm is 96.38 and is greater than K-means (77.25), Y-means (87.15) and FCM (82.13). Also false alarm rate is 3.2 which is lesser than that of K-means (5.2), Y-means (3.9) and FCM (4.2).

The Table 2 shows the percentage of various attacks including denial of service attack percentage, remote to local attack percentage, user to root attack percentage and probe attack percentage which are used as a measuring parameter to check the percentage of accuracy and detection rate of our proposed algorithm and various existing algorithms.

Table 2.Experiment Results of Proposed Algorithm, K-means, FCM, and Y-means

| Parameters | K-means | FCM | Y-means | Proposed algorithm |
|---|---|---|---|---|
| Denial of service (%) | 79.83 | 83.12 | 89.15 | 96.12 |
| Remote to local (%) | 78.12 | 82.45 | 85.10 | 90.10 |
| User to root (%) | 52.10 | 60.10 | 65.12 | 70.51 |
| Probe (%) | 62.45 | 65.25 | 68.12 | 70.13 |

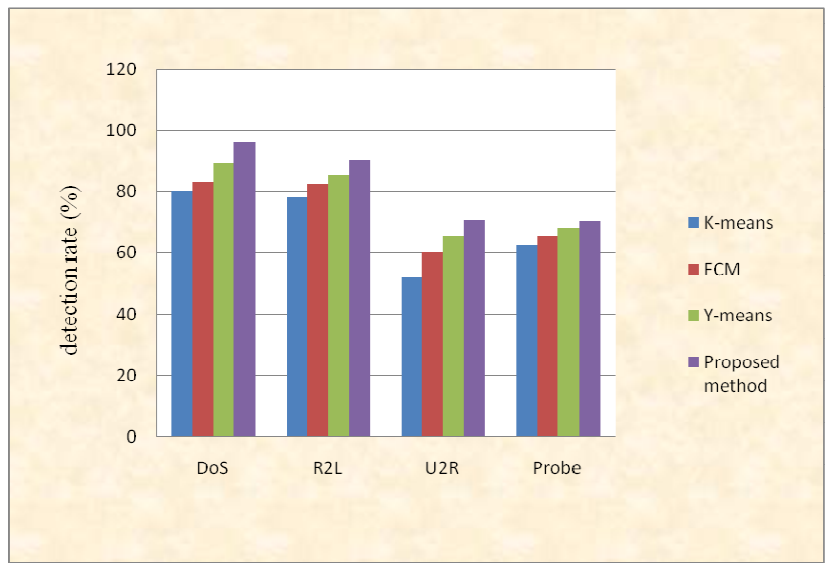

Figure 3. Comparisons of Detection Rate against various attacks

The figure 3 illustrates the percentage of detection rate of denial of service attack, remote to local attack, user to root attack and probe attack and comparisons are made for proposed method, K-means algorithm, Y-means algorithm and FCM algorithm. We observed from the figure that detection rate for our proposed approach is greater than earlier approaches.

The result findings show that the proposed algorithm is able to detect greater number of intrusions with higher accuracy compared to existing algorithm. The detection rate of Denial of service attack and Remote to local attack is greater than user to Root attack and Probe attack. Results also show that there are lesser false positive rate and false negative rate.

In order to improve detection rate and accuracy a modified version of k-medoid algorithm is used and this eliminates the disadvantages of K-means algorithm. Also the degeneracy is being eliminated by using the concept of searching for the empty clusters and deleting them.

The initial medoid selection is the important task for the algorithm since it affects the performance of the proposed algorithm. By using an efficient method for the selection of initial medoid the algorithm performs faster execution and overcomes the disadvantages of existing algorithm.

## 6. CONCLUSION AND FUTURE WORK

The approach proposed in this paper describes the new way of intrusion detection using k-medoid clustering algorithm and certain modifications of it. The algorithm specified a new way of selection of initial medoid and proved to be better than K-means for anomaly intrusion detection. The algorithm conveys the idea of data mining technologies which is certainly a good field and popular area of research in intrusion detection. The proposed approach is having many advantages over the existing algorithm which mainly overcomes the disadvantages of dependency on initial centroids, dependency on the number of cluster and irrelevant clusters. The algorithm is able to sort out these problems and has been able to provide high detection rates and less false negative rate. The algorithm has many advantages but there are few disadvantages which need to be focused. The detection rate can for probe and user to root attack can be further enhanced by efficient method of clustering which is our future work.

## ACKNOWLEDGEMENTS

The authors wish to thank the faculty members of department of Information Technology,B.I.T. Mesra, Ranchi, for contributing a lot of effort in providing with resources, study materials and reading manuscripts and for devoting their valuable time for this paper and thanks are also due to A.K. Akhtar for his suggestions on preparing the manuscript.

**AUTHORS**

Ravi Ranjan is pursuing his M.Tech in Information Security from Birla Institute of Technology, Mesra, Ranchi, India since 2012 and going to complete in the year 2014. He received his B.E in Information science and engineering in the year 2011. His research interest includes computer networks, data mining, cloud computing, information security, compiler design.

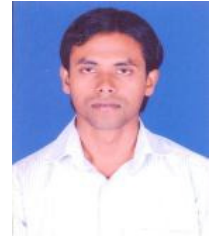

G. Sahoo received his MSc in Mathematics from Utkal University in the year 1980 and PhD in the Area of Computational Mathematics from Indian Institute of Technology, Kharagpur in the year 1987. He has been associated with Birla Institute of Technology, Mesra, Ranchi, India since 1988, and currently, he is working as a Professor and Head in the Department of Information Technology. His research interest includes theoretical computer science, parallel and distributed computing, cloud computing, evolutionary computing, information security, image processing and pattern recognition.

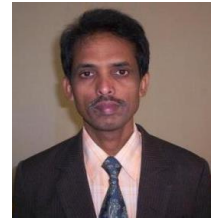